\documentclass[a4paper]{article}
\RequirePackage[english]{babel}
\RequirePackage[latin1]{inputenc}
\RequirePackage[T1]{fontenc}
\RequirePackage{mathrsfs}
\RequirePackage{amsmath}
\RequirePackage{amssymb}
\RequirePackage{amsbsy}
\RequirePackage{bm}
\begin{document}
\title{\huge{\bf{Causal propagation for ELKO fields}}}
\author{Luca Fabbri\\ 
\footnotesize (Gruppo di Fisica Teorica, INFN \& University of Bologna, Italy)}
\date{}
\maketitle
\begin{abstract}
We shall consider the general problem of causal propagation for spinor fields, focus attention in particular on the case constituted by ELKO fields and will show that the problem of causal propagation for ELKO fields is always solvable.
\end{abstract}
\section*{Introduction}
The problem of the causal propagation of matter fields, first discussed by Velo and Zwanziger in \cite{v-z}, shows that matter fields may have bad propagation if higher-spins are considered: the reason for this fact is that higher-spin fields are postulated to satisfy higher-spin field equations, whose special forms, in presence of gauge interactions, let the aforementioned problems occur. Nevertheless, least-spin fields can still have bad propagation if they obey higher-order differential equations, because in these field equations the second-order derivatives of the field are derivatives of the torsion of the field, whose form gives rise to non-linearities and complicated dynamics leading to the same kind of problems.

On the other hand, the ELKO fields, recently introduced by Ahluwalia and Grumiller in \cite{a-g/1} and \cite{a-g/2}, are defined to be least-spin spinor fields described by higher-order differential equations, that is precisely the form of matter for which causality problems may arise.

In this paper, we will analyze the causal propagation for ELKO fields to see whether it can always be ensured.
\section{Causal propagation}
In the following, we will consider the causality problem as exposed by Velo and Zwanziger in their original paper \cite{v-z}, and hereby we will briefly recall the general concepts. Roughly speaking, the analysis of Velo and Zwanziger is based on this fact: the equation that determines the propagation of the wavefronts is obtained by considering in the field equations only the highest-order derivative terms, then formally replacing the derivatives with a vector $n$ and requiring the resulting equation to be singular; thus done, one obtains an equation called characteristic equation, whose solutions are in terms of the vector $n$ and such that if $n$ has the temporal component larger than the euclidean norm of the spatial components, then the fields have superluminal propagation.

In the simplest case in which the spinorial field verifies the simplest spinorial field equations, the characteristic equation is
\begin{eqnarray}
\mathrm{det}(\gamma^{\mu}n_{\mu})=0
\label{equation-least-least}
\end{eqnarray}
which gives $n$ with null norm maintaining the field within the light-cone; otherwise, to escape this situation it is necessary for the characteristic equation to be more complicated, which is the case when the field equations contain more terms of the highest-order derivatives: this situation occurs whenever higher-spin fields are considered to be governed by higher-spin field equations, whose special forms give characteristic equations of general structure
\begin{eqnarray}
\mathrm{det}(\gamma^{\mu}n_{\mu}g_{\nu\alpha}
+\Gamma_{\alpha}n_{\nu}+G_{\nu}n_{\alpha})=0
\label{equation-first-high}
\end{eqnarray}
in terms of some matrix $\Gamma_{\alpha}$ and $G_{\nu}$, for which the vector $n$ can have positive norms and thus the field can be boosted out of the light-cone. However, also for least-spin spinor fields there can be problems if they are governed by higher-order differential equations, since in these cases the second-order derivatives of the field become derivatives of the torsion of the field, and this yields characteristic equations of the general structure
\begin{eqnarray}
\mathrm{det}(An^{2}+C^{\mu\nu}n_{\nu}n_{\mu})=0
\label{equation-second}
\end{eqnarray}
in terms of some spinorial matrix $A$ and $C^{\mu\nu}$, admitting possible positive norms for $n$ that correspond to propagation out of the light-cone for the fields.
\section{ELKO fields}
\subsection{ELKO fields: structure of differential equations}
We shall now consider the theory of ELKOs as exposed by Ahluwalia and Grumiller in their original papers \cite{a-g/1} and \cite{a-g/2}, and here we recall the general ideas. The ELKOs are spinors with spin-$\frac{1}{2}$ transformation law defined to be eigenstates of the charge-conjugation operator $\lambda=\gamma^{2}\lambda^{*}$ up to a complex phase, as explained in the original references above and in \cite{r-r} in terms of their classification, and as it is also explained in \cite{r-h} where the relationship between ELKOs and Dirac spinors is discussed; because of this definition ELKOs enjoy special features, the most important of which being that they have mass dimension equal to $1$ so that they must be described by second-order derivative Lagrangians, as it is explained in the references mentioned above and also in \cite{h-r} where the relationship between ELKO and Dirac Lagrangians is studied. Finally, as it is carried along in \cite{b-b} and \cite{b/1}, the ELKOs and their dynamics can be generalized to the torsional case.

For these spinors the first property that needs be highlighted is that a generic ELKO $\lambda$ has its own definition of ELKO dual $\stackrel{\neg}{\lambda}$, as it is shown and justified in the aforementioned references; for them the derivatives $D_{\mu}$ are defined as usual, and in terms of the contorsionless derivatives $\nabla_{\mu}$ we have the decomposition 
\begin{eqnarray}
&D_{\mu}\lambda
=\nabla_{\mu}\lambda
+\frac{1}{2}K^{ij}_{\phantom{ij}\mu}\sigma_{ij}\lambda\\
&D_{\mu}\stackrel{\neg}{\lambda}
=\nabla_{\mu}\stackrel{\neg}{\lambda}
-\frac{1}{2}\stackrel{\neg}{\lambda}\sigma_{ij}K^{ij}_{\phantom{ij}\mu}
\label{derivatives}
\end{eqnarray}
in terms of the contorsion tensor $K^{\rho}_{\phantom{\rho}\mu\nu}$ and depending on the $\sigma$ matrices, whereas the commutator of derivatives 
\begin{eqnarray}
&[D_{\mu},D_{\nu}]\lambda
=Q^{\rho}_{\phantom{\rho}\mu\nu}D_{\rho}\lambda
+\frac{1}{2}G^{ij}_{\phantom{ij}\mu\nu}\sigma_{ij}\lambda\\
&[D_{\mu},D_{\nu}]\stackrel{\neg}{\lambda}
=Q^{\rho}_{\phantom{\rho}\mu\nu}D_{\rho}\stackrel{\neg}{\lambda}
-\frac{1}{2}\stackrel{\neg}{\lambda}\sigma_{ij}G^{ij}_{\phantom{ij}\mu\nu}
\label{commutator}
\end{eqnarray}
is defined in terms of the torsion tensor $Q^{\rho}_{\phantom{\rho}\mu\nu}$ and the curvature tensor $G^{\rho}_{\phantom{\rho}\eta\mu\nu}$ and it depends on the $\sigma$ matrices as well: both the contorsion and torsion tensors have one independent contraction given by $Q^{\rho}_{\phantom{\rho}\rho\nu}=K_{\nu\rho}^{\phantom{\rho\nu}\rho}=K_{\nu}=Q_{\nu}$ which we shall call Cartan vector, whereas the curvature tensor has only one independent contraction given by $G^{\rho}_{\phantom{\rho}\eta\rho\nu}=G_{\eta\nu}$ which we shall call Ricci tensor, with only one contraction given by $G_{\eta\nu}g^{\eta\nu}=G$ called Ricci scalar. Finally, by taking the product of two derivatives of the spinor and the Ricci scalar one builds general second-order derivative Lagrangians.

We postulate the Lagrangian
\begin{eqnarray}
L=G+D_{\mu}\stackrel{\neg}{\lambda}D^{\mu}\lambda-m^{2}\stackrel{\neg}{\lambda}\lambda
\label{internal-lagrangian}
\end{eqnarray}
in terms of $m$ representing the mass of the field, and being the only parameter of the model.

By varying this Lagrangian with respect to the spinor field, we get the corresponding field equations for the spinor as
\begin{eqnarray}
D^{2}\lambda+K^{\mu}D_{\mu}\lambda+m^{2}\lambda=0
\label{equationspinorfield}
\end{eqnarray}
which are second-order derivative field equations, with derivatives that contain contorsion and for which contorsion can be separated by writing
\begin{eqnarray}
\nabla^{2}\lambda+\frac{1}{2}\nabla_{\mu}K^{\alpha\beta\mu}\sigma_{\alpha\beta}\lambda
+K^{ij\mu}\sigma_{ij}\nabla_{\mu}\lambda+\frac{1}{4}K^{ij\mu}K_{ab\mu}\sigma_{ij}\sigma^{ab}\lambda
+m^{2}\lambda=0
\label{equationspinor}
\end{eqnarray}
identically.

By varying the Lagrangian with respect to any connection, or with respect to the contorsion, we get the corresponding field equations for the spin as 
\begin{equation}
(K_{\mu[\alpha\beta]}+K_{[\alpha}g_{\beta]\mu})=
\frac{1}{2}\left(D_{\mu}\stackrel{\neg}{\lambda}\sigma_{\alpha\beta}\lambda
-\stackrel{\neg}{\lambda}\sigma_{\alpha\beta}D_{\mu}\lambda\right)
\label{equationsspin}
\end{equation}
which relate the contorsion tensor to the spin tensor, this last tensor being written in terms of the spinor field derivatives in which contorsion can be separated apart by writing them as
\begin{eqnarray}
\nonumber
&(K_{\mu[\alpha\beta]}+K_{[\alpha}g_{\beta]\mu})=\\
&=\frac{1}{2}(\nabla_{\mu}\stackrel{\neg}{\lambda}\sigma_{\alpha\beta}\lambda
-\stackrel{\neg}{\lambda}\sigma_{\alpha\beta}\nabla_{\mu}\lambda)
-\frac{1}{4}K_{\sigma\rho\mu}
\stackrel{\neg}{\lambda}\{\sigma_{\alpha\beta},\sigma^{\sigma\rho}\}\lambda
\label{equations}
\end{eqnarray}
and in which by means of the properties of the $\sigma$ matrices we get to the form
\begin{eqnarray}
\nonumber
&4(K_{\mu\alpha\beta}-K_{\mu\beta\alpha}+K_{\alpha}g_{\beta\mu}-K_{\beta}g_{\alpha\mu})
+\frac{1}{2}K^{\sigma\rho}_{\phantom{\sigma\rho}\mu}
\varepsilon_{\alpha\beta\sigma\rho}(i\stackrel{\neg}{\lambda}\gamma\lambda)-\\
&-K_{\alpha\beta\mu}(\stackrel{\neg}{\lambda}\lambda)
-2(\nabla_{\mu}\stackrel{\neg}{\lambda}\sigma_{\alpha\beta}\lambda
-\stackrel{\neg}{\lambda}\sigma_{\alpha\beta}\nabla_{\mu}\lambda)=0
\label{relationship}
\end{eqnarray}
where in the mixed term the product of the bilinear spinors and the contorsion tensor results to be factorized in order to let us reach a considerably simpler form for this relationship; however, this relationship is such that it determines contorsion as an implicit function of the contorsionless derivatives of the spinor field, and in order to invert it so as to explicitly get contorsion in terms of the contorsionless derivatives of the spinor field itself we may proceed by writing it as
\begin{eqnarray}
\nonumber
&4(K_{\mu\alpha\beta}+K_{\beta\mu\alpha}+K_{\alpha\beta\mu})
+4(K_{\alpha}g_{\beta\mu}-K_{\beta}g_{\alpha\mu})
+\frac{1}{2}K^{\sigma\rho}_{\phantom{\sigma\rho}\mu}
\varepsilon_{\alpha\beta\sigma\rho}(i\stackrel{\neg}{\lambda}\gamma\lambda)-\\
&-K_{\alpha\beta\mu}(4+\stackrel{\neg}{\lambda}\lambda)
-2(\nabla_{\mu}\stackrel{\neg}{\lambda}\sigma_{\alpha\beta}\lambda
-\stackrel{\neg}{\lambda}\sigma_{\alpha\beta}\nabla_{\mu}\lambda)=0
\end{eqnarray}
in which the first term is the completely antisymmetric part of contorsion to be written as
\begin{eqnarray}
\nonumber
&2(K_{\nu\theta\eta}\varepsilon^{\nu\theta\eta\rho})\varepsilon_{\mu\alpha\beta\rho}
+4(K_{\alpha}g_{\beta\mu}-K_{\beta}g_{\alpha\mu})
+\frac{1}{2}K^{\sigma\rho}_{\phantom{\sigma\rho}\mu}
\varepsilon_{\alpha\beta\sigma\rho}(i\stackrel{\neg}{\lambda}\gamma\lambda)-\\
&-K_{\alpha\beta\mu}(4+\stackrel{\neg}{\lambda}\lambda)
-2(\nabla_{\mu}\stackrel{\neg}{\lambda}\sigma_{\alpha\beta}\lambda
-\stackrel{\neg}{\lambda}\sigma_{\alpha\beta}\nabla_{\mu}\lambda)=0
\label{intermediate}
\end{eqnarray}
and upon multiplication by $\varepsilon^{\alpha\beta\zeta\xi}$ we first get
\begin{eqnarray}
\nonumber
&4(K_{\nu\theta\eta}\varepsilon^{\nu\theta\eta\xi}g^{\mu\zeta}
-K_{\nu\theta\eta}\varepsilon^{\nu\theta\eta\zeta}g^{\mu\xi})
+8\varepsilon^{\alpha\mu\zeta\xi}K_{\alpha}
+2K^{\zeta\xi\mu}(i\stackrel{\neg}{\lambda}\gamma\lambda)-\\
&-K_{\alpha\beta}^{\phantom{\alpha\beta}\mu}\varepsilon^{\alpha\beta\zeta\xi}
(4+\stackrel{\neg}{\lambda}\lambda)
-2(\nabla^{\mu}\stackrel{\neg}{\lambda}\sigma_{\alpha\beta}\lambda
-\stackrel{\neg}{\lambda}\sigma_{\alpha\beta}\nabla^{\mu}\lambda)\varepsilon^{\alpha\beta\zeta\xi}=0
\label{auxiliary}
\end{eqnarray}
whose contraction is 
\begin{equation}
2K^{\xi}(i\stackrel{\neg}{\lambda}\gamma\lambda)
-K_{\alpha\beta\zeta}\varepsilon^{\alpha\beta\zeta\xi}
(8-\stackrel{\neg}{\lambda}\lambda)
+2(\nabla_{\zeta}\stackrel{\neg}{\lambda}\sigma_{\alpha\beta}\lambda
-\stackrel{\neg}{\lambda}\sigma_{\alpha\beta}\nabla_{\zeta}\lambda)\varepsilon^{\alpha\beta\zeta\xi}=0
\end{equation}
while coming back to (\ref{intermediate}) we see that its internal contraction is given by
\begin{equation}
\frac{1}{2}K^{\sigma\rho\mu}\varepsilon_{\sigma\rho\mu\beta}
(i\stackrel{\neg}{\lambda}\gamma\lambda)
-K_{\beta}(8-\stackrel{\neg}{\lambda}\lambda)
-2(\nabla^{\mu}\stackrel{\neg}{\lambda}\sigma_{\mu\beta}\lambda
-\stackrel{\neg}{\lambda}\sigma_{\mu\beta}\nabla^{\mu}\lambda)=0
\end{equation}
and these last two contractions give us a linear system of two equations in which the trace and the completely antisymmetric parts of the contorsion are the only two independent variables, so that we can solve the linear system for the two variables to get
\begin{eqnarray}
\nonumber
&K^{\xi}
\left[(8-\stackrel{\neg}{\lambda}\lambda)^{2}
-(i\stackrel{\neg}{\lambda}\gamma\lambda)^{2}\right]=\\
\nonumber
&=2(8-\stackrel{\neg}{\lambda}\lambda)
(\nabla_{\mu}\stackrel{\neg}{\lambda}\sigma^{\xi\mu}\lambda
-\stackrel{\neg}{\lambda}\sigma^{\xi\mu}\nabla_{\mu}\lambda)+\\
&+(i\stackrel{\neg}{\lambda}\gamma\lambda)
(\nabla_{\zeta}\stackrel{\neg}{\lambda}\sigma_{\alpha\beta}\lambda
-\stackrel{\neg}{\lambda}\sigma_{\alpha\beta}\nabla_{\zeta}\lambda)\varepsilon^{\alpha\beta\zeta\xi}
\label{trace}
\end{eqnarray}
and
\begin{eqnarray}
\nonumber
&K_{\alpha\beta\zeta}\varepsilon^{\alpha\beta\zeta\xi}
\left[(8-\stackrel{\neg}{\lambda}\lambda)^{2}
-(i\stackrel{\neg}{\lambda}\gamma\lambda)^{2}\right]=\\
\nonumber
&=4(i\stackrel{\neg}{\lambda}\gamma\lambda)
(\nabla_{\mu}\stackrel{\neg}{\lambda}\sigma^{\xi\mu}\lambda
-\stackrel{\neg}{\lambda}\sigma^{\xi\mu}\nabla_{\mu}\lambda)+\\
&+2(8-\stackrel{\neg}{\lambda}\lambda)
(\nabla_{\zeta}\stackrel{\neg}{\lambda}\sigma_{\alpha\beta}\lambda
-\stackrel{\neg}{\lambda}\sigma_{\alpha\beta}\nabla_{\zeta}\lambda)\varepsilon^{\alpha\beta\zeta\xi}
\label{completelyantisymmetric}
\end{eqnarray}
giving the trace and the completely antisymmetric parts of the contorsion itself, which can eventually be used in equation (\ref{auxiliary}) to work out from equation (\ref{intermediate}) the final result
\begin{eqnarray}
\nonumber
&K_{\alpha\beta\mu}
\left[(8-\stackrel{\neg}{\lambda}\lambda)^{2}
-(i\stackrel{\neg}{\lambda}\gamma\lambda)^{2}\right]
\left[(4+\stackrel{\neg}{\lambda}\lambda)^{2}
-(i\stackrel{\neg}{\lambda}\gamma\lambda)^{2}\right]=\\
\nonumber
&=\left[(8-\stackrel{\neg}{\lambda}\lambda)^{2}
-(i\stackrel{\neg}{\lambda}\gamma\lambda)^{2}\right]
(i\stackrel{\neg}{\lambda}\gamma\lambda)
(\stackrel{\neg}{\lambda}\sigma^{\sigma\rho}\nabla_{\mu}\lambda
-\nabla_{\mu}\stackrel{\neg}{\lambda}\sigma^{\sigma\rho}\lambda)
\varepsilon_{\sigma\rho\alpha\beta}+\\
\nonumber
&+2\left[(8-\stackrel{\neg}{\lambda}\lambda)^{2}
-(i\stackrel{\neg}{\lambda}\gamma\lambda)^{2}\right]
(4+\stackrel{\neg}{\lambda}\lambda)
(\stackrel{\neg}{\lambda}\sigma_{\alpha\beta}\nabla_{\mu}\lambda
-\nabla_{\mu}\stackrel{\neg}{\lambda}\sigma_{\alpha\beta}\lambda)-\\
\nonumber
&-4\left[(4+\stackrel{\neg}{\lambda}\lambda)
(8-\stackrel{\neg}{\lambda}\lambda)
-(i\stackrel{\neg}{\lambda}\gamma\lambda)^{2}\right]
(\stackrel{\neg}{\lambda}\sigma_{\sigma\theta}\nabla_{\zeta}\lambda
-\nabla_{\zeta}\stackrel{\neg}{\lambda}\sigma_{\sigma\theta}\lambda)
\varepsilon_{\alpha\beta\mu\rho}\varepsilon^{\sigma\theta\zeta\rho}+\\
\nonumber
&+8\left[(4+\stackrel{\neg}{\lambda}\lambda)
(8-\stackrel{\neg}{\lambda}\lambda)
-(i\stackrel{\neg}{\lambda}\gamma\lambda)^{2}\right]
(\stackrel{\neg}{\lambda}\sigma_{\eta\alpha}\nabla^{\eta}\lambda
-\nabla^{\eta}\stackrel{\neg}{\lambda}\sigma_{\eta\alpha}\lambda)g_{\mu\beta}-\\
\nonumber
&-8\left[(4+\stackrel{\neg}{\lambda}\lambda)
(8-\stackrel{\neg}{\lambda}\lambda)
-(i\stackrel{\neg}{\lambda}\gamma\lambda)^{2}\right]
(\stackrel{\neg}{\lambda}\sigma_{\eta\beta}\nabla^{\eta}\lambda
-\nabla^{\eta}\stackrel{\neg}{\lambda}\sigma_{\eta\beta}\lambda)g_{\mu\alpha}-\\
\nonumber
&-16(2-\stackrel{\neg}{\lambda}\lambda)
(i\stackrel{\neg}{\lambda}\gamma\lambda)
(\stackrel{\neg}{\lambda}\sigma^{\eta\rho}\nabla_{\eta}\lambda
-\nabla_{\eta}\stackrel{\neg}{\lambda}\sigma^{\eta\rho}\lambda)\varepsilon_{\alpha\beta\mu\rho}+\\
\nonumber
&+8(2-\stackrel{\neg}{\lambda}\lambda)
(i\stackrel{\neg}{\lambda}\gamma\lambda)
(\stackrel{\neg}{\lambda}\sigma^{\sigma\theta}\nabla^{\zeta}\lambda
-\nabla^{\zeta}\stackrel{\neg}{\lambda}\sigma^{\sigma\theta}\lambda)
g_{\mu\beta}\varepsilon_{\sigma\theta\zeta\alpha}-\\
&-8(2-\stackrel{\neg}{\lambda}\lambda)
(i\stackrel{\neg}{\lambda}\gamma\lambda)
(\stackrel{\neg}{\lambda}\sigma^{\sigma\theta}\nabla^{\zeta}\lambda
-\nabla^{\zeta}\stackrel{\neg}{\lambda}\sigma^{\sigma\theta}\lambda)
g_{\mu\alpha}\varepsilon_{\sigma\theta\zeta\beta}
\label{contorsion}
\end{eqnarray}
where contorsion is expressed in terms of the contorsionless derivatives of the field identically.

This expression can be inserted into equations (\ref{equationspinor}) where the contorsionless derivatives of contorsion of the field is the contorsionless derivatives of contorsionless derivatives of the field, resulting in field equations containing only contorsionless derivatives of the field but such that the highest-order contorsionless derivatives of the field has now got a more complicated structure. 

And as a consequence of the more complicated structure these field equations will have, we shall expect a correspondingly complicated structure of the characteristic equation. 
\subsection{ELKO fields: causally propagating solutions}
The resulting characteristic equation is indeed complicated, as it is of the general form given by (\ref{equation-second}) where
\begin{eqnarray}
\nonumber
&A=\left[(8-\stackrel{\neg}{\lambda}\lambda)^{2}
-(i\stackrel{\neg}{\lambda}\gamma\lambda)^{2}\right]
\left[(4+\stackrel{\neg}{\lambda}\lambda)^{2}
-(i\stackrel{\neg}{\lambda}\gamma\lambda)^{2}\right]\mathbb{I}+\\
\nonumber
&+\left[(8-\stackrel{\neg}{\lambda}\lambda)
(4-\stackrel{\neg}{\lambda}\lambda)(4+\stackrel{\neg}{\lambda}\lambda)
-(i\stackrel{\neg}{\lambda}\gamma\lambda)^{2}
(\stackrel{\neg}{\lambda}\lambda)\right]
\sigma_{\alpha\beta}\lambda\stackrel{\neg}{\lambda}\sigma^{\alpha\beta}+\\
&+\left[(8-\stackrel{\neg}{\lambda}\lambda)^{2}
-(i\stackrel{\neg}{\lambda}\gamma\lambda)^{2}\right]
\left(\frac{i\stackrel{\neg}{\lambda}\gamma\lambda}{2}\right)
\sigma_{\alpha\beta}\lambda\stackrel{\neg}{\lambda}\sigma_{\sigma\rho}\varepsilon^{\sigma\rho\alpha\beta}
\end{eqnarray}
and
\begin{eqnarray}
C^{\mu}_{\eta}=
8(2-\stackrel{\neg}{\lambda}\lambda)(i\stackrel{\neg}{\lambda}\gamma\lambda)
(\sigma_{\eta\rho}\lambda\stackrel{\neg}{\lambda}\sigma_{\sigma\theta}
+\sigma_{\sigma\theta}\lambda\stackrel{\neg}{\lambda}\sigma_{\eta\rho})\varepsilon^{\sigma\theta\rho\mu}
\end{eqnarray}
with a structure that can admit for $n$ time-like solutions.

However, it is clear that causally propagating solutions do exist; to see that consider first that contorsion can actually be expressed in terms of the fields only if some bilinear spinors are limited by bounds given by weak-field conditions: since these weak-field conditions can be considered in the most extreme case where the bilinear spinors are all much smaller than the unity, the above matrices are such that $C^{\mu\eta}$ is negligible compared to $A$ which can be developed in series reducing to
\begin{eqnarray}
A\approx\mathbb{I}+\frac{1}{4}\stackrel{\neg}{\lambda}\lambda\mathbb{I}
+\frac{1}{8}\sigma_{\alpha\beta}\lambda\stackrel{\neg}{\lambda}\sigma^{\alpha\beta}
\end{eqnarray}
implying that the characteristic equation (\ref{equation-second}) admits only light-like solutions for $n$ corresponding to causal propagation for the fields.

This indicates that under extreme weak-field conditions causal propagation for these spinors is ensured, even though these extreme weak-field conditions are not needed to ensure causal propagation should particular forms of the ELKOs be assumed.
\section*{Conclusion}
In the present paper, we have shown that the causal propagation for ELKO fields can always be preserved.

Because causal propagation for ELKO fields is assured, these field theories are clearly strengthened.

Furthermore, it has been recently pointed out in \cite{a-l-s} that the ELKO structure endows the theory with a preferred axis of locality; this is in line with the existence of a preferred direction whose evidence has been discussed in \cite{l-m}.

Moreover, ELKO models appear to be one of the most promising form of matter that can account for inflation, as in \cite{s}, in \cite{b-m}, \cite{b/11} and \cite{b/2}. And finally, ELKO theories seems to be one of the best candidate for dark matter, as presented in \cite{a} and \cite{D}.

All these aspects of ELKO theories certainly contribute to let them occupy a privileged place within the standard model of cosmology.

\

\noindent \textbf{Acknowledgments.} I am grateful to Professor Giorgio Velo and Christian G.~B\"{o}hmer for the enlightening discussions on the subject.

\



\begin{thebibliography}{10}
\bibitem{v-z}
G.~Velo and D.~Zwanziger,
\textit{Phys. Rev.} \textbf{186}, 1337 (1969).
\bibitem{a-g/1}
D.~V.~Ahluwalia and D.~Grumiller,
JCAP \textbf{0507}, 012 (2005).
\bibitem{a-g/2}
D.~V.~Ahluwalia and D.~Grumiller,
\textit{Phys. Rev. D} \textbf{72}, 067701 (2005).
\bibitem{r-r}
R.~da Rocha and W.~A.~J.~Rodrigues,
\textit{Mod. Phys. Lett. A} \textbf{21}, 65 (2006).
\bibitem{r-h}
R.~da Rocha and J.~M.~Hoff da Silva,
\textit{J. Math. Phys.} \textbf{48}, 123517 (2007).
\bibitem{h-r}
J.~M.~Hoff da Silva and R.~da Rocha,
\textit{Int. J. Mod. Phys. A} \textbf{24}, 3227 (2009).
\bibitem{b-b}
C.~G.~B\"{o}hmer and J.~Burnett,
\textit{Phys. Rev. D} \textbf{78}, 104001 (2008).
\bibitem{b/1}
C.~G.~B\"{o}hmer,
\textit{Annalen Phys.} \textbf{16}, 38 (2007).
\bibitem{a-l-s}
D.~V.~Ahluwalia, C.~Y.~Lee and D.~Schritt,
arXiv: 0911.2947 [hep-ph].
\bibitem{l-m}
K.~Land and J.~Magueijo,
\textit{Mon. Not. Roy. Astron. Soc.} \textbf{378}, 153 (2007).
\bibitem{s}
S.~Shankaranarayanan,
arXiv: 0905.2573 [astro-ph.CO].
\bibitem{b-m}
C.~G.~B\"{o}hmer and D.~F.~Mota,
\textit{Phys. Lett. B} \textbf{663}, 168 (2008).
\bibitem{b/11}
C.~G.~Boehmer,
\textit{Annalen Phys.} \textbf{16}, 325 (2007).
\bibitem{b/2}
C.~G.~Boehmer,
\textit{Phys. Rev. D} \textbf{77}, 123535 (2008).
\bibitem{a}
D.~V.~Ahluwalia,
\textit{Int. J. Mod. Phys. D} \textbf{15}, 2267 (2006).
\bibitem{D}
DAMA Collaboration,
\textit{Mod. Phys. Lett. A} \textbf{23}, 2125 (2008).
\end{thebibliography}
\end{document}